\documentclass[aps,prl,twocolumn,floatfix,superscriptaddress,preprintnumbers]{revtex4-1}

\pdfoutput=1
\usepackage{amsmath,amssymb}
\usepackage{amsfonts}
\usepackage{bm,bbm}
\usepackage{graphicx}
\usepackage{mathrsfs}
\usepackage{slashed}
\usepackage{booktabs}
\usepackage{tabu}
\usepackage[dvipsnames]{xcolor}
\usepackage[normalem]{ulem}
\usepackage{tikz}
\usepackage{soul}
\usepackage[colorlinks=true,linkcolor=blue]{hyperref}
\usepackage{xcolor}
\usepackage{ulem}
\usepackage{array}
\usepackage{verbatim}
\usepackage{epsfig}
\usepackage{multirow}

\newcommand{\nsrcA}{n_{\rm {\scriptscriptstyle SRC}}^A}
\newcommand{\nsrcD}{n_{\rm {\scriptscriptstyle SRC}}^d}
\newcommand{\REMCd}{R_{\rm {\scriptscriptstyle EMC}}^d}
\newcommand{\pF}{p_{\!\scriptscriptstyle F}^{}}

\begin{document}

\preprint{ADP-20-11/T1121, JLAB-THY-20-3173}

\title{Do short-range correlations cause the nuclear EMC effect in the deuteron?}

\author{X.~G.~Wang}
\affiliation{CSSM and ARC Centre of Excellence for Particle Physics at the Terascale, Department of Physics, University of Adelaide SA 5005 Australia}
\author{A.~W.~Thomas}
\affiliation{CSSM and ARC Centre of Excellence for Particle Physics at the Terascale, Department of Physics, University of Adelaide SA 5005 Australia}
\author{W. Melnitchouk}
\affiliation{Jefferson Lab, Newport News, Virginia 23606, USA \\}

\begin{abstract}
The relative contributions to the valence nuclear EMC effect in the deuteron arising from off-shell effects and Fermi motion are examined in models which include nuclear binding and off-shell effects. Contrary to expectations, the effect of Fermi motion overwhelms the off-shell effects for nucleons in short-range correlations (SRCs), calling into question the hypothesized causal connection between SRCs and the EMC effect.
\end{abstract}

\date{\today}
\maketitle

%%%%%%%%%%%%%%%%%%%%%%%%%%%%%%%%%%%%%%%%%%%%%%%%%%%%%%%%%%%%%%%%%%%%%%%
The discovery by the European Muon Collaboration (EMC) of the unexpected suppression of the deep-inelastic structure functions of atomic nuclei in the valence quark region~\cite{Aubert:1983xm, Ashman:1988bf, Benvenuti:1987az, Gomez:1993ri}, which we shall refer to as the ``valence EMC effect'', still has no widely accepted explanation after more than three decades~\cite{Geesaman:1995yd, Norton:2003cb, Bickerstaff:1989ch, Berger:1987er}.
Experimental advances have yielded precise data across the periodic table~\cite{Fomin:2011ng, Seely09}, while many theoretical ideas have been proposed, ranging from nucleon ``swelling''~\cite{Close83, Close84}, to enhancement of the pion field~\cite{LlewellynSmith83, Ericson83, Berger84}, off-shell effects~\cite{Dunne86}, suppression of point-like configurations~\cite{Frankfurt88} and modification of nucleon structure in the strong scalar and vector mean fields
occurring in nuclei~\cite{Thomas:1989vt, Saito:1992rm, Smith02, Mineo:2003vc, Cloet:2006bq}.
In a few cases new predictions have been made against which some of these ideas may be tested in future experiments~\cite{Cloet:2012td, Thomas:2018kcx}, but to date a universally accepted explanation has remained stubbornly elusive.

In the past few years there has been remarkable progress at Jefferson Lab in clearly identifying the tensor force as the prime source of short-range correlations (SRCs) in nuclei~\cite{Duer:2018sby}. Following that success, the observation of a correlation between the size of the valence EMC effect and the number of nucleons in SRCs led to suggestions that SRCs may be the underlying source of the effect~\cite{Schmookler19}.
The physical motivation is that in deep-inelastic scattering (DIS) involving nucleons in SRCs, the spectator to the DIS event will have a relatively high kinetic energy, which forces the struck nucleon to be far off-mass-shell.
Within this picture it was shown that one could extract an approximately ``universal function'', applicable to all nuclei including the deuteron, which describes the modification of the structure function of such a far off-shell nucleon.

In this Letter we examine the proposal that SRCs may be responsible for the valence EMC effect in more detail.
In order to place the theoretical discussion on a quantitative level we use a general class of phenomenologically successful models of DIS from nuclei, which take into account nuclear binding, Fermi motion, and nucleon off-shell effects.
The key feature of these models is that, irrespective of their details, they allow these effects to be separated into those arising from low- and high-momentum nucleons. 
The results suggest that the effect of Fermi motion is extremely important if one is to draw conclusions regarding the role of SRCs in the valence EMC effect.

Schmookler {\it et al.}~\cite{Schmookler19} assumed that the structure function of nucleus $A$ could be decomposed into contributions from unmodified mean-field protons and neutrons and contributions from $np$ pairs in SRCs with modified structure functions,
\begin{eqnarray}
\label{eq:F2A}
F_2^A &=& ( Z - \nsrcA ) F_2^p + ( N - \nsrcA ) F_2^n  + \nsrcA (F_2^{p*} + F_2^{n*}) \nonumber\\
&=& Z F_2^p + N F_2^n + \nsrcA (\Delta F_2^p + \Delta F_2^n)\, ,
\end{eqnarray}
where $\nsrcA$ is the number of $np$ pairs in the nucleus $A$, and the structure functions depend on the Bjorken $x$ variable and the exchanged four-momentum squared, $Q^2$.
The off-shell nucleon structure modifications are denoted by
\begin{equation}
\label{eq:DeltaF2N}
\Delta F_2^N = F_2^{N*} - F_2^N, \ \ \ \ N=p, n ,
\end{equation}
where $F_2^{N*}$ is the averaged modified structure function for nucleons in SRC pairs.
Inherent in this formulation is the assumption that all of the modifications of the nucleon structure can be absorbed in the form of the off-shell corrections, $\Delta F_2^N$.
The relative modification of the structure functions of nucleons in $np$ SRC pairs,
\begin{equation}
\dfrac{\nsrcD (\Delta F_2^p + \Delta F_2^n)}{F_2^d} 
= \dfrac{\dfrac{F_2^A}{F_2^d} 
- (Z - N) \dfrac{F_2^p}{F_2^d} - N}{(A/2) a_2 - N},
\end{equation}
where $a_2 = (2/A) (\nsrcA/\nsrcD)$, was extracted from the EMC data as a universal function for all nuclei.
Applying Eq.~(\ref{eq:F2A}) to a deuteron target, this universal function is related to the deuteron structure function by
\begin{equation}
\label{eq:R}
R\, \equiv\, \dfrac{F_2^d - (F_2^p + F_2^n)}{F_2^d} 
= \dfrac{\nsrcD (\Delta F_2^p + \Delta F_2^n)}{F_2^d} \, .
\end{equation}

For comparison with the phenomenological analysis of Ref.~\cite{Schmookler19}, we consider the nuclear structure function within the relativistic impulse approximation, in which the nuclear DIS takes place via incoherent scattering from bound, off-shell nucleons~\cite{Gross92, MST94, MST94plb, KPW94, Afnan03, Pace01, Sargsian02, Rinat03, KP06, KP10, Tropiano19}. 
One can show that in this case the nuclear $F_2^A$ structure function can be written as a sum of a convolution term involving on-shell nucleon structure functions and an off-shell correction,
\begin{equation}
\label{eq:F2d}
F_2^d(x) = \big( f_{p/d} \otimes F_2^p\, +\, f_{n/d} \otimes F_2^n \big)(x)\, +\, F_2^{d (\rm off)}(x) \, ,
\end{equation}
where the convolution symbol $\otimes$ is defined by the integral
    $(f \otimes F_2)(x) = \int_x^A dy f(y)\, F_2(x/y)$ 
and for simplicity we have suppressed the $Q^2$ dependence in the functions.
The smearing function $f_{p/d} = f_{n/d} \equiv f_{N/d}$ describes the distribution of nucleons in the deuteron carrying a fraction $y$ of the deuteron's (light-cone) momentum,
\begin{equation}
\label{eq:fy}
f_{N/d}(y) = \int\!dp^2\, \widetilde{f}_{N/d}(y,p^2),
\end{equation}
where $p^2 = p_0^2 - \bm{p}^2$ the invariant mass squared of the off-shell nucleon, and can be straightforwardly computed from the deuteron wave function,
\begin{equation}
\label{eq:fyp2}
\widetilde{f}_{N/d}(y,p^2) \propto \left| \psi_d(p) \right|^2.
\end{equation}
On the other hand, the off-shell term, $F_2^{d (\rm off)}$,  isolates the dependence on the medium modification of the nucleon structure and is more model dependent.
One should caution that the off-shell term $F_2^{d (\rm off)}$ in Eq.~(\ref{eq:F2d}) (as well as the off-shell term $\Delta F_2^N$ in Eq.~(\ref{eq:DeltaF2N})) is not physical, as one can in principle move strength between the different terms in Eq.~(\ref{eq:F2d}).
Nevertheless, in the literature a number of attempts have been made to estimate the nucleon off-shell modifications consistently within specific models.

Working within a covariant % hybrid
quark--hadron framework, Melnitchouk {\it et al.}~\cite{MST94, MST94plb} computed $F_2^d$ using relativistic nucleon--deuteron vertex functions~\cite{Buck79} (see also~\cite{Gross07, Gross10}).
These include $P$-wave admixtures and take into account Lorentz scalar and vector interactions in the deuteron. 
For the nucleon itself this approach used a spectator quark model parametrizing off-shell nucleon--quark--spectator vertex functions for spin-0 and spin-1 di-quark spectators.
The off-shell correction, $F_2^{d (\rm off)}$, was given by a sum of non-factorized terms depending on the $P$-state wave functions and the nucleon virtuality
    $v(p^2) \equiv (p^2 - M^2)/M^2$,
where $M$ is the nucleon mass.
Integrating over all $p^2$, the off-shell effects in this model were found to be $\lesssim 1\%\!-\!2\%$ in the relevant range of $x$~\cite{MST94plb}.

A similar formulation by Kulagin {\it et al.}~\cite{KPW94, KP06, KMPW95, KM08d} expands the off-shell nucleon scattering amplitude in powers of the nucleon momentum $\bm{p}$, allowing the nuclear structure function to be expressed as a generalized convolution of the nuclear spectral function and a $p^2$-dependent nucleon structure function, $\widetilde{F}_2^N$, % with corrections of ${\cal O}(v^2)$,
\begin{equation}
\label{eq:F2ful}
F_2^d(x) = \sum_N \int\!dy\, dp^2\, \widetilde{f}_{N/d}(y,p^2)\, \widetilde{F}_2^N(x/y,p^2).
\end{equation}
Expanding the off-shell nucleon function in a Taylor series around $v=0$ one finds
\begin{equation}
\label{eq:F2Noff}
\widetilde{F}_2^N(x,p^2)
= F_2^N(x) \Big( 1 + v(p^2)\, \delta\!f(x) + {\cal O}(v^2) \Big)
\, .
\end{equation}
The off-shell correction term $F_2^{d (\rm off)}$ in Eq.~(\ref{eq:F2d}) can then be written as a convolution of an off-shell smearing function, $f_{N/d}^{(\rm off)}$, and the nucleon off-shell function, $\delta f$~\cite{Tropiano19, Ehlers14}
\begin{equation}
\label{eq:F2doff}
F_2^{d (\rm off)}(x) = \sum_N \Big( f_{N/d}^{(\rm off)} \otimes \big[ F_2^N \cdot \delta f \big] \Big)(x),
\end{equation}
where
\begin{equation}
\label{eq:fNdoff}
f_{N/d}^{(\rm off)}(y) = \int\!dp^2\ v(p^2)\, \widetilde{f}_{N/d}(y,p^2) \, .
\end{equation}
Assuming that the off-shell structure function $\widetilde{F}_2^N$ has a spectral representation, the off-shell dependence of the quark spectral function was modelled in terms of the $p^2$ dependence of the ultraviolet cutoff parameter regularizing the integration over the quark momentum~\cite{KP06, Ethier13}, which it was argued could be related to the confinement radius and the amount of nucleon swelling in the nuclear medium~\cite{CJ11}.
Alternatively, rather than rely on models to estimate the off-shell function $\delta\!f$, several analyses have more recently sought to parametrize the function phenomenologically and determine the parameters directly from a global QCD analysis of high energy proton and deuteron data~\cite{KP10, CJ15}.

\begin{figure}[t]
\includegraphics[width=0.95\columnwidth]{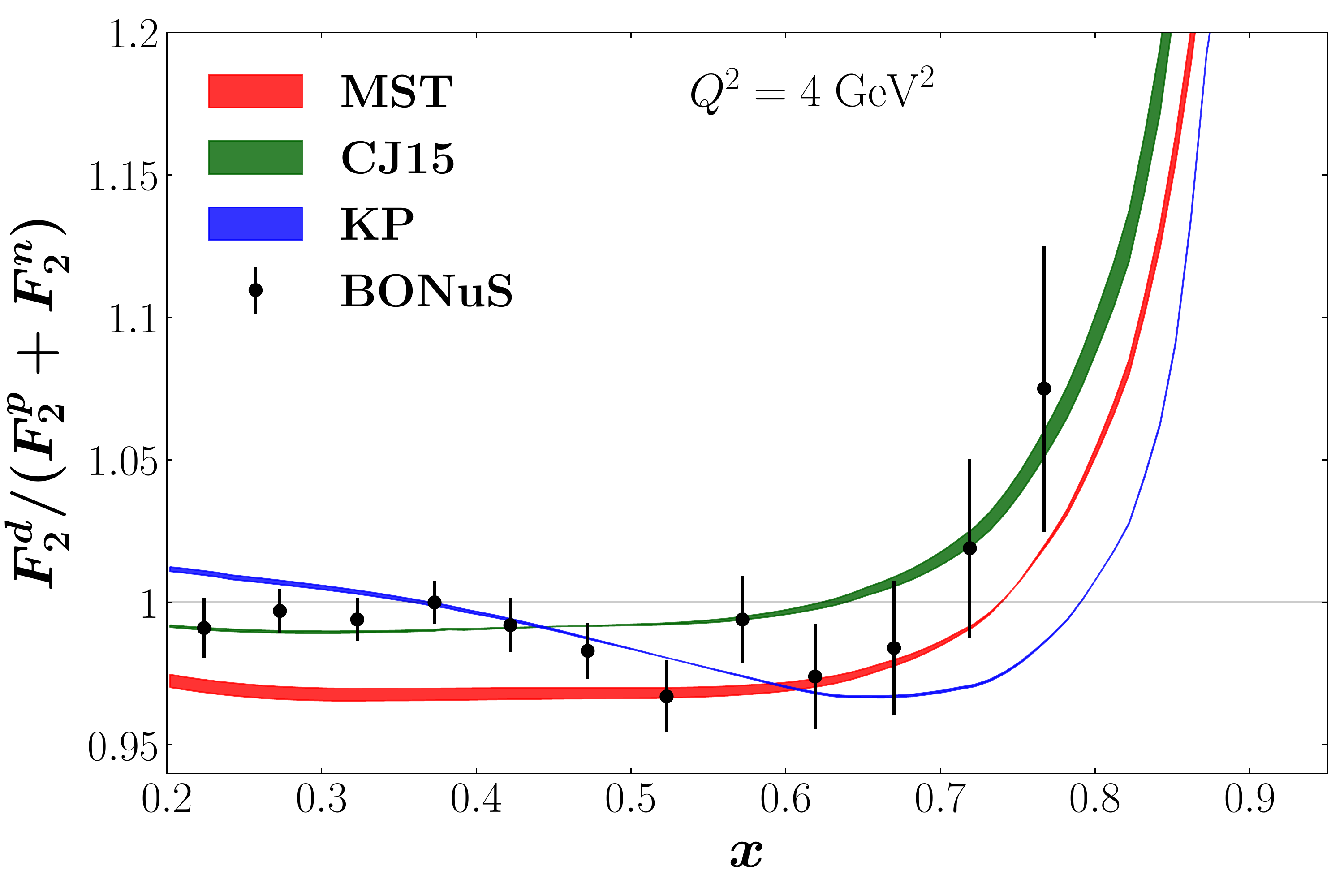}
\caption{Ratio of deuteron to isoscalar nucleon structure functions $F_2^d/(F_2^p+F_2^n)$ from the relativistic spectator model (MST) calculation \cite{MST94plb} (red band), the CJ15 global QCD analysis \cite{CJ15} (green band), and the Kulagin-Petti (KP) phenomenological analysis \cite{KP06} (blue band), compared with the data from the BONuS experiment \cite{BONuS15} (black circles). The bands represent results using the Paris~\cite{D-Paris} and \mbox{WJC-2}~\cite{D-WJC} deuteron wave functions.}
\label{fig:F2dN}
\end{figure}

The results of these analyses for the nuclear EMC ratio in the deuteron,
\begin{equation}
\label{eq:F2dN}
\REMCd \equiv \frac{F_2^d}{F_2^p+F_2^n}\,
=\, \frac{1}{1-R}\, \approx\, 1 + R\ \ \ {\rm for}\ R \ll 1,
\end{equation}
are illustrated in Fig.~\ref{fig:F2dN} at a fixed $Q^2 = 4$~GeV$^2$, and compared with data from the BONuS experiment at Jefferson Lab~\cite{BONuS12, BONuS14, BONuS15}.
This experiment measured tagged low-momentum protons in coincidence with scattered electrons in DIS from the deuteron, with $Q^2$ ranging from 1.2~GeV$^2$ at the lowest $x$ value shown up to 4.0~GeV$^2$ at the highest $x$.
Within the uncertainties quoted by the experiment (in Fig.~\ref{fig:F2dN} the statistical and systematic uncertainties have been added in quadrature), the inclusive data cannot at present definitively discriminate between the different off-shell behaviors in the models.

The essential difference for our purposes between all the models~\cite{Schmookler19, MST94, MST94plb, KP06, CJ11, CJ15} discussed above is the relative amount of the valence EMC effect that is attributed to the off-shell effects and especially to off-shell effects associated with nucleons in SRCs. 
Whereas in the convolution models of Eq.~(\ref{eq:F2d}) the off-shell effects are just one source of difference between the free nucleon and nuclear structure functions, the model in Eq.~(\ref{eq:F2A}) represents the most extreme scenario whereby {\it all} of the nuclear corrections are attributed to off-shell nucleon modification for nucleons in SRCs, Eq.~(\ref{eq:DeltaF2N}).

\begin{figure}[t]
\includegraphics[width=0.95\columnwidth]{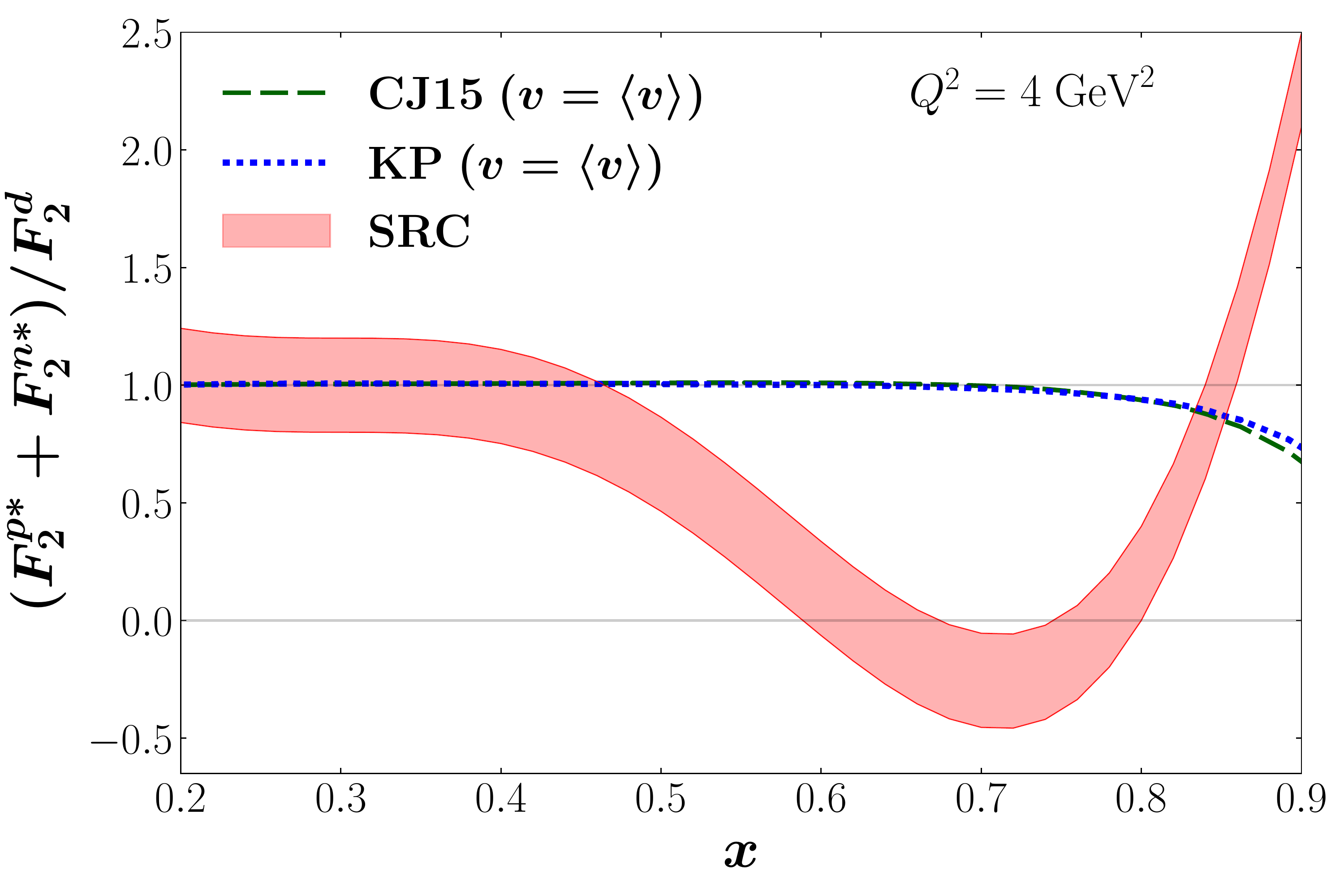}
\caption{Ratio of off-shell nucleon structure function to the deuteron structure function, $(F_2^{p*} + F_2^{n*})/F_2^d$, for the models discussed in Fig.~\ref{fig:F2dN}. The ratio for the SRC model is computed from Eq.~(\ref{eq:F2offSRC}) (the red band represents the uncertainty on the SRC parametrization extracted from Ref.~\cite{Schmookler19}), while for the CJ15 and KP models the off-shell nucleon functions are computed from Eq.~(\ref{eq:F2Noff}) at an average nucleon virtuality \mbox{$v = \langle v \rangle$}.}
\label{fig:F2off}
\end{figure}

This is more dramatically illustrated in Fig.~\ref{fig:F2off}, where we show the ratio of the (unphysical) off-shell nucleon structure function to the deuteron structure function, $(F_2^{p*} + F_2^{n*})/F_2^d$.
For the SRC model, from Eqs.~(\ref{eq:DeltaF2N}) and (\ref{eq:R}) this can be computed from the ratio $R$ and the number $\nsrcD$ of $np$ pairs in the deuteron,
\begin{equation}
\label{eq:F2offSRC}
\frac{F_2^{p*} + F_2^{n*}}{F_2^d} = 1 + \frac{R}{\nsrcD}(1 - \nsrcD).
\end{equation}
The probability of finding nucleons in an SRC pair in the deuteron can be evaluated by integrating the deuteron wave function over large relative momenta,
\begin{equation}
\nsrcD = \int_{\pF}^{\infty} d|{\bm{p}}|\, \bm{p}^2 \big| \psi_d(p) \big|^2.
\end{equation}
Taking a typical choice for the demarcation between low momentum nucleons and those in SRCs to be at the Fermi momentum $p_F = 300$~MeV, one finds $\nsrcD = 3.4\%$   
for the Paris deuteron wave function~\cite{D-Paris}, while 
for the relativistic WJC-2 deuteron wave function~\cite{D-WJC} one has $\nsrcD = 3.8\%$.
The resulting ratio is close to unity at low~$x$, but then decreases dramatically to even go negative at $x \sim 0.6-0.8$, before increasing again at higher $x$ values to mimic the effect of Fermi motion.

In contrast, the off-shell nucleon structure functions in the convolution approach appear under the integral in Eq.~(\ref{eq:F2ful}).
Defining the average value of the virtuality $v$~by
\begin{equation}
\langle v \rangle = \int dy\, f_{N/d}^{(\rm off)}(y),
\end{equation}
we find
    $\langle v \rangle = -4.3\%$    % -4.34\%$
for the Paris wave function and
    \mbox{$-5.0\%$}                 % -4.99\% 
for the WJC-2 wave function.
The ratio in Fig.~\ref{fig:F2off} can then be evaluated according to
\begin{equation}
\label{eq:F2offCONV}
\frac{F_2^{p*} + F_2^{n*}}{F_2^d}
= \frac{\big\langle \widetilde{F}_2^p + \widetilde{F}_2^n \big\rangle}{F_2^d}
= \frac{(F_2^p + F_2^n) (1 + \delta\!f\, \langle v \rangle)}{F_2^d} \, .
\end{equation}
Since the off-shell function $\delta f$ is generally $\lesssim {\cal O}(1)$, the ratio (\ref{eq:F2offCONV}) deviates from unity only very slightly due to the small value of $\langle v \rangle$.
In fact, the ratios in Fig.~\ref{fig:F2off} for the convolution models are basically the inverse of the EMC effect ratios shown in Fig.~\ref{fig:F2dN}, with deviations from this proportional to $\delta\!f\, \langle v \rangle$ that are negligible.

While not ruled out phenomenologically by the inclusive structure function measurements, the interpretation of a negative off-shell nucleon structure function in the SRC model, which appears already for $x \gtrsim 0.6$, is problematic.
The striking contrast between the two pictures illustrated in Fig.~\ref{fig:F2off} suggests the need to further explore whether the off-shell scenarios can be distinguished in specific regions of $p^2$, such as those typically associated with SRCs.

In fact, the formulation of the convolution models in Eqs.~(\ref{eq:F2d}) and (\ref{eq:F2ful}) explicitly in terms of the momentum dependent functions makes it possible to separate the contributions to the deuteron structure function from the low- and high-momentum regions,
\begin{equation}
F_2^d = F_2^d\big|_{p < \pF} + F_2^d\big|_{p > \pF} \, ,
\label{eq:F2dpF}
\end{equation}
where the first and second terms represent contributions to the integrals in Eqs.~(\ref{eq:fy}), (\ref{eq:F2ful}) and (\ref{eq:fNdoff}) from below and above the Fermi momentum, $p_F$.
If SRCs are indeed responsible for the observed valence EMC effect, as hypothesized in Ref.~\cite{Schmookler19}, one should find that the $p > p_F$ term in Eq.~(\ref{eq:F2dpF}) plays a leading role in generating the differences between the nuclear and free-nucleon structure functions in the valence region.
\begin{widetext}

\begin{figure}[t]
\includegraphics[width=\columnwidth]{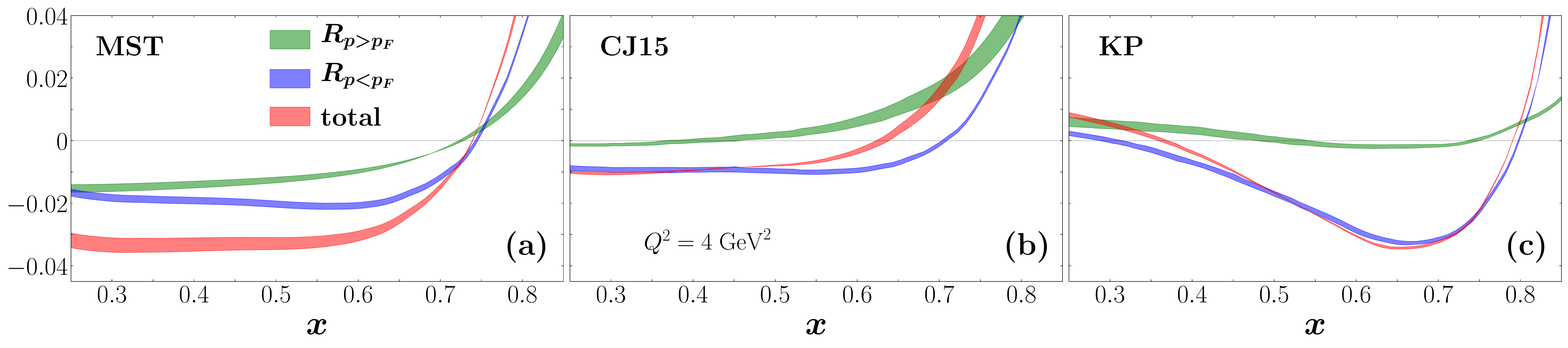}
\caption{Contributions to the ratio $R = \big(F_2^d-(F_2^p+F_2^n)\big)/F_2^d$ from the regions $p>p_F$ (green bands), $p<p_F$ (blue bands), and the total (red bands), for the
    {\bf (a)} MST~\cite{MST94plb},
    {\bf (b)} CJ15~\cite{CJ15}, and 
    {\bf (c)} KP~\cite{KP06} models
for the nucleon off-shell corrections at $Q^2=4$~GeV$^2$. The bands envelop the results obtained using the nonrelativistic Paris~\cite{D-Paris} and the relativistic WJC-2~\cite{D-WJC} deuteron wave functions.}
\label{fig:RpF}
\end{figure}

\end{widetext}
To test this hypothesis, we use Eq.~(\ref{eq:F2dpF}) to decompose the ratio $R$ in Eq.~(\ref{eq:R}) into low-momentum and high-momentum components,
\begin{equation}
R = R_{p>\pF} + R_{p<\pF} \, ,
\end{equation}
where
\begin{subequations}
\label{eq:RpF}
\begin{eqnarray}
\label{eq:Rp>pF}
R_{p>\pF} &=& \dfrac{F_2^d\big|_{p>\pF} - \nsrcD (F_2^p + F_2^n)}{F_2^d} ,\\
\label{eq:Rp<pF}
R_{p<\pF} &=& \dfrac{F_2^d\big|_{p<\pF} - (1 - \nsrcD) (F_2^p + F_2^n)}{F_2^d} \, .
\end{eqnarray}
\end{subequations}
The individual contributions $R_{p>\pF}$ and $R_{p<\pF}$, as well as the total ratio $R$, are shown in Fig.~\ref{fig:RpF} for each of the three microscopic models considered~\cite{MST94plb, KP06, CJ15}.
As may be expected from Eq.~(\ref{eq:F2dN}), for each model the shape of the total $R$ ratio follows that of the $\REMCd$ ratio in Fig.~\ref{fig:F2dN}.
On the other hand, the relative contributions to the ratio from the low- and high-momentum regions reveal features that are not evident in the integrated quantities.

In particular, for the MST model~\cite{MST94plb} in Fig.~\ref{fig:RpF}(a) one finds that in the valence EMC effect region, $0.3 \lesssim x \lesssim 0.7$, less than half of the strength comes from the high-momentum $p>p_F$ region at the lower $x$ values, and even less at the higher-$x$ end of the range.
For even larger $x$ values, $x \to 1$, the rise in $R$ due to Fermi motion is also more strongly associated with the $p<p_F$ region.

For the CJ15 case in Fig.~\ref{fig:RpF}(b), the overall nuclear EMC effect is found to be the smallest, as also evident from Fig.~\ref{fig:F2dN}, with $\approx 1/3$ of the magnitude of that in the MST model.
However, within the small absolute size of the total effect, essentially 100\% comes from the the $p<p_F$ region for $x \lesssim 0.5$, while at larger $x$ values, $x \sim 0.6-0.7$, the low- and high-$p$ contributions largely cancel. 
In contrast, for the KP model in Fig.~\ref{fig:RpF}(c), where the EMC effect follows the general characteristics of the effect in heavy nuclei, the deviation from zero is greatest for $x \approx 0.6-0.7$, where it is completely dominated by the low-momentum,  $p<p_F$, contributions.

The inescapable conclusion to be drawn from the results in Fig.~\ref{fig:RpF} is that, even though the shapes and magnitudes of the EMC effect differ considerably between the different models and analyses, in none of the cases is the effect dominated by the high-momentum $p>p_F$ region associated with SRCs.
The total valence EMC effect is either dominated by the low-momentum $p<p_F$ components (as in the KP model and in the CJ15 model at lower $x$) or received approximately similar contributions from both regions (as in MST model and in the CJ15 case at higher $x$).
This finding is independent of the choice of deuteron wave function, with essentially the same results obtained for the widely-used Paris wave function~\cite{D-Paris} and the more recent relativistic WJC-2 wave functions~\cite{D-WJC}.

We have also investigated the effect of the variation of the $\bm{p}$ boundary, in particular, when the Fermi momentum is raised to $p_F = 400$~MeV.
In this case the number of nucleons in SRCs decreases slightly to $\nsrcD = 2.0\%$ for the Paris wave function and $2.6\%$ for the WJC-2 wave function.
Compared with results obtained with $p_F=300$~MeV as the boundary for SRCs, the magnitude of the $p<p_F$ contribution increases for the larger $p_F$ and becomes closer to the total result in each case in Fig.~\ref{fig:RpF}.
This can be expected from the definition of $R_{p>\pF}$ and $R_{p<\pF}$ in Eqs.~(\ref{eq:RpF}):
when $p_F \to 0$, $R_{p>\pF} \to R$ and $R_{p<\pF} \to 0$, while
when $p_F \to \infty$, $R_{p>\pF} \to 0$ and $R_{p<\pF} \to R$.

Clearly this analysis does not support the hypothesis that there is a causal connection between nucleons residing in SRCs and the EMC effect.
Instead, it suggests that the effect in the deuteron is dominated by binding and to a lesser extent off-shell effects associated with nucleons in the region $p<p_F$. 
It is crucial to this conclusion that neglecting the effect of Fermi motion is a particularly poor approximation, especially for nucleons in SRCs, and in no model do SRCs give more than half the EMC effect in deuterium.

Future avenues to explore the origin of the nuclear EMC effect include semi-inclusive DIS on the deuteron and other light nuclei, with tagging of recoil protons and neutrons to leverage the virtuality of the scattered nucleon~\cite{MSS97}.
Asymmetric nuclei can provide additional information on the isospin dependence of the off-shell effects, which has been raised in recent studies of DIS from $^3$He~\cite{Tropiano19}.
Finally, there are dramatically different predictions for the spin \cite{Cloet:2006bq, Thomas:2018kcx} and flavor \cite{Cloet:2012td} dependence of the EMC effect in microscopic models and those predicted from the SRC picture.
Data from upcoming experiments at Jefferson Lab or the future Electron-Ion Collider could be instrumental in finally resolving the long sought-after origin of the nuclear EMC effect.

%%%%%%%%%%%%%%%%%%%%%%%%%%%%%%%%%%%%%%%%%%%%%%%%%%%%%%%%%%%%%%%%%%%%%%%%
\vspace{0.3cm}
This work was supported by the Australian Research Council through the Discovery Projects DP151103101 and DP180100497 (AWT) and the US Department of Energy Contract No.~DE-AC05-06OR23177, under which Jefferson Science Associates, LLC operates Jefferson Lab (WM).

%%%%%%%%%%%%%%%%%%%%%%%%%%%%%%%%%%%%%%%%%%%%%%%%%%%%%%%%%%%%%%%%%%%%%%%%

\end{document}